\documentclass[a4paper,10pt]{article}
\usepackage{amsmath,amssymb}
\begin{document}

\title{\textbf{SLAVNOV-TAYLOR IDENTITY FOR NONEQUILIBRIUM QUARK-GLUON PLASMA}}
\author{K. OKANO\footnote{E-mail: okano@sci.osaka-cu.ac.jp}}
\date{\footnotesize{\textit{Department of Physics, Osaka City University, Sumiyoshi-ku, Osaka 558-8585, Japan}}}
\maketitle
\begin{abstract}
\footnotesize{Within the closed-time-path formalism of nonequilibrium QCD, we derive a Slavnov-Taylor (ST) identity for the gluon polarization
 tensor. The ST identity takes the same form both in Coulomb and
 covariant gauges. Application to quasi-uniform quark-gluon plasma (QGP)
 near equilibrium or nonequilibrium quasistationary QGP is made.}
\end{abstract}

\section{Introduction}
 Much interest is devoted to the physics of a deconfinement phase of
 hadronic matter (quark-gluon plasma, QGP), with both strong experimental
 and theoretical research going on. A theoretical understanding of
 this new phase of matter can be gained in the framework of hot QCD
 supplemented with a perturbative approach and important progress has
 been made during the last decade [1]. At early stages, the QGP is
 treated as a thermally and/or chemically equilibrium system. Studies of
 the QGP as a nonequilibrium system have recently begun. 

 The symmetry transformation which leaves the Lagrangian invariant plays
 an important role. Each continuous symmetry leads to relations between
 Green functions. As an example, we take up the BRST invariance of the
 QCD Lagrangian, which leads to Slavnov-Taylor (ST) identities. In a
 previous work [2], we derived a ST identity for gluon polarization
 tensor in equilibrium QGP within the imaginary-time formalism [3,4]. As
 an application of it, we dealt with damping rates for soft gluons. The purpose of this letter is to derive a ST identity for
 the case of nonequilibrium QGP. We employ the closed-time-path (CTP)
 formalism [5,6,7] of nonequilibrium QCD. It turns out that the deduced ST identity takes the same form both in Coulomb and covariant gauges.
 
\section{Preliminaries}
We start with the QCD Lagrangian density in Coulomb gauge:
\begin{eqnarray}
\mathcal{L}(A,\omega, \bar{\omega})&=&-\frac{1}{4}F_{\mu\nu}^{a}(x)F^{a,\mu\nu}(x)-\frac{\lambda}{2}(\boldsymbol{\nabla\cdot A}^{a}(x))(\boldsymbol{\nabla\cdot A}^{a}(x))\nonumber \\
& &-\partial^{\underline{\mu}}\bar{\omega}^{a}(x)D_{\underline{\mu}}^{ac}(A(x))\omega^{c}(x),
\end{eqnarray}
\begin{equation}
F_{\mu\nu}^{a}(x)=\partial_{\mu}A_{\nu}^{a}(x)-\partial_{\nu}A_{\mu}^{a}(x)+gf^{abc}A_{\mu}^{b}(x)A_{\nu}^{c}(x),
\end{equation}
where $a, b, c$ are the color indices, 
$D_{\mu}^{ac}(A(x))\equiv\delta^{ac}\partial /\partial
x^{\mu}+gf^{abc}A_{\mu}^{b}(x)$, and
$x^{\underline{\mu}}\equiv x^{\mu}-x_{0}n^{\mu}$ with $n^{\mu}=(1,\boldsymbol{0})$. $\omega^{a} $ and $ \bar{\omega}^{a}$ are the FP-ghost fields (in
Coulomb gauge). The quark sector does not play any role for our purpose. 

The CTP formalism is constructed by introducing an oriented closed-time
path $C$ $(=C_{1}\oplus C_{2})$ in a complex-time plane, that goes from
$-\infty$ to $+\infty$ $(C_{1})$ and then returns from $+\infty$ to
$-\infty$ $(C_{2})$. The time arguments of the fields are on the time path
$C$. A field with time argument on $C_{1}$ $[C_{2}]$ is called a
type-1 [type-2] field. A classical contour action is written in the form
\begin{eqnarray}
& &\int_{C}dx_{0}\int d^{3}x\mathcal{L}(A,\omega ,\bar{\omega})\nonumber \\
& &~~~~~~~~=\int_{-\infty}^{+\infty}dx_{0}\int d^{3}x\Bigl[\mathcal{L}(A_{1},\omega_{1},\bar{\omega}_{1})-\mathcal{L}(A_{2},\omega_{2},\bar{\omega}_{2})\Bigr]\nonumber \\
& &~~~~~~~~\equiv\int_{-\infty}^{+\infty}dx_{0}\int d^{3}x\hat{\mathcal{L}}(x),
\end{eqnarray}
where the subscripts ``1'' and ``2'' stand for the type of fields and $\hat{\mathcal{L}}$
is sometimes called a hat-Lagrangian [8]. 

The (full) gluon propagator $G_{rs}^{ab,\mu\nu}(x,y)$ is defined by the statistical average of the
time-path-ordered product $T_{C}$ of gluon fields:   
\begin{equation}
G_{rs}^{ab,\mu\nu}(x,y)\equiv\mbox{Tr}\Bigl[T_{C}(A_{r}^{a,\mu}(x)A_{s}^{b,\nu}(y))\rho\Bigr]\equiv\langle T_{C}(A_{r}^{a,\mu}(x)A_{s}^{b,\nu}(y))\rangle .
\end{equation}
Here $\rho$ is the density matrix [7]:
\begin{equation}
\rho =\int\mathcal{D}A_{\mu}^{a}(\boldsymbol{x})\mathcal{D}A_{\nu}^{\prime b}(\boldsymbol{x})|A_{\mu}^{a}(\boldsymbol{x})\rangle\rho(A_{\mu}^{a},A_{\nu}^{\prime b})\langle A_{\nu}^{\prime b}(\boldsymbol{x})|,
\end{equation}
where $|A_{\mu}^{a}(\boldsymbol{x})\rangle$ is the eigenstate of the
in-field operator
$A_{\mu}^{a}(x_{0}=-\infty,\boldsymbol{x})$.\footnote{In the case
of covariant gauge, the FP-ghost fields should also be included.} The (full) ghost propagator
$\tilde{G}_{rs}^{ab}(x,y)$ reads
\begin{equation}
\tilde{G}_{rs}^{ab}(x,y)\equiv\langle T_{C}(\omega_{r}^{a}(x)\bar{\omega}_{s}^{b}(y))\rangle.
\end{equation}
The bare ghost propagator $\tilde{\Delta}_{rs}^{ab}(x,y)$ is constructed
from Eq.(3) with Eq.(1):
\begin{equation}
\tilde{\Delta}_{rs}^{ab}(x,y)=\delta^{ab}(-1)^{r-1}\delta_{rs}\int\frac{d^{4}P}{(2\pi)^{4}}\,\frac{i}{p^{2}}~e^{-iP\cdot(x-y)}
\end{equation}
with no summation over $r$, $P^{\mu}=(p_{0}, \boldsymbol{p})$, and $p\equiv|\boldsymbol{p}|$.
The gluon-ghost vertex factor for the ``type-1'' and ``type-2'' vertices can be read off from \\$\hat{\mathcal{L}}\ni
(-1)^{t}gf^{abc}\partial^{\underline{\mu}}\omega_{t}^{a}A_{t,\underline{\mu}}^{b}\omega_{t}^{c}$:
\begin{equation}
\mathcal{V}_{t}=(-1)^{t-1}gf^{abc}P^{\underline{\mu}}~~~~(t=1,2),
\end{equation}
where $P^{\underline{\mu}}=P^{\mu}-p_{0}n^{\mu} (=(0,\boldsymbol{p}))$ with $\boldsymbol{p}$ the momentum of the out-going FP-ghost, and
`$\underline{\mu}$' is a suffix of $A_{t,\underline{\mu}}^{b}$. 

We introduce here the generating functional:
\begin{eqnarray}
 \hat{Z}[J,\bar{\xi},\xi]&=&\int\prod_{r=1}^{2}\mathcal{D}A_{r,\mu}^{a}\mathcal{D}\omega_{r}^{a}\mathcal{D}\bar{\omega}_{r}^{a}\nonumber \\
& &\times\exp\biggl[i\int_{-\infty}^{+\infty}d^4z \Bigl\{\hat{\mathcal{L}}+\sum_{t=1}^{2}(J_{t}^{a,\mu}A_{t,\mu}^{a}+\bar{\xi}_{t}^{a}\omega_{t}^{a}\nonumber \\
& &+\bar{\omega}_{t}^{a}\xi_{t}^{a})\Bigr\}\biggr]\rho(A_{\mu}^{a},A_{\nu}^{\prime b}),
\end{eqnarray}
where $ J, \bar{\xi}, \xi$ are (classical) source functions. Equation (9) is to be computed with periodic boundary
conditions, $A_{1,\mu}^{a}(x_{0}=-\infty,
\boldsymbol{x})=A_{2,\mu}^{a}(x_{0}=-\infty, \boldsymbol{x})$,
$\omega_{1}^{a}(x_{0}=-\infty, \boldsymbol{x})=\omega_{2}^{a}(x_{0}=-\infty,
\boldsymbol{x})$ and $\bar{\omega}_{1}^{a}(x_{0}=-\infty,
\boldsymbol{x})=\bar{\omega}_{2}^{a}(x_{0}=-\infty, \boldsymbol{x})$. These
conditions come from the trace operation [4,6]. Note that, inspite of the
fact that the  ghost fields are fermionic, they
obey a periodic boundary condition [4,9]. As is stated above after
Eq.(5), $A_{\mu}^{a}$ and $A_{\nu}^{\prime b}$ of $\rho$ in Eq.(9) are the
eigenvalues of the in-fields, i.e., the fields at $x_{0}=-\infty$, so
that $A_{\mu}^{a}=A_{1,\mu}^{a}=A_{2,\mu}^{a}$ etc. $\hat{Z}[J,\bar{\xi},\xi]$ generates the above defined (full) propagators through 
\begin{eqnarray}
G_{rs}^{ab,\mu\nu}(x,y)&=&\frac{\delta\ln\hat{Z}[J,\bar{\xi},\xi]}{i\delta J_{s,\nu}^{b}(y)i\delta J_{r,\mu}^{a}(x)}\Bigg|_{J=\bar{\xi}=\xi=0},\\
\tilde{G}_{rs}^{ab}(x,y)&=&\frac{\delta\ln\hat{Z}[J,\bar{\xi},\xi]}{i\delta\xi_{s}^{b}(y)i\delta\bar{\xi}_{r}^{a}(x)}\Bigg|_{J=\bar{\xi}=\xi=0}.
\end{eqnarray}

The Lagrangian density $\mathcal{L}$ is invariant [10] under the BRST transformation:
\begin{eqnarray}
& &\delta A_{\mu}^{a}=\zeta D_{\mu}^{ac}(A)\omega^{c} ,~~~
\delta\omega^{a}=-\frac{1}{2}g\zeta f^{abc}\omega^{b}\omega^{c},\nonumber \\
& &\delta\bar{\omega}^{a}=\lambda\zeta\boldsymbol{\nabla\cdot A}^{a},
\end{eqnarray}
where $\zeta$ is a Grassmann-number parameter. Throughout in the sequel,
we deal with the systems whose density matrix $\rho$ is invariant under
the BRST transformation.
Using these facts for Eq.(9), we obtain 
\begin{eqnarray}
& &\int d^{4}z\mathcal{B}(z)\hat{Z}[J,\bar{\xi},\xi]=0,\nonumber \\
& &\mathcal{B}(z)=\sum_{t=1}^{2}\Biggl[J_{t}^{a,\mu}(z)D_{\mu}^{ac}\Bigl(\frac{\delta}{i\delta J_{t}(z)}\Bigr)\frac{\delta}{i\delta\bar{\xi}_{t}^{c}(z)} +\lambda\xi_{t}^{a}(z)\frac{\partial}{\partial z_{\underline{\mu}}}\frac{\delta}{i\delta J_{t}^{a,\underline{\mu}}(z)}\nonumber \\
& &~~~~~~~~~~~~~~~~~~+\bar{\xi}_{t}^{a}(z)\frac{g}{2}f^{abc}\frac{\delta}{i\delta\bar{\xi}_{t}^{b}(z)}\frac{\delta}{i\delta\bar{\xi}_{t}^{c}(z)}\Biggr].
\end{eqnarray}

\section{Slavnov-Taylor identity}
In the following we deal with systems, for which $ \langle
A_{r}^{a,\mu}(x)\rangle =0 $ holds. Computing
\begin{equation}
\frac{\delta}{i\delta J_{r,\mu}^{a}(x)}\frac{\delta}{i\delta\xi_{s}^{b}(y)}\int d^{4}z\mathcal{B}(z)\ln\hat{Z}[J,\bar{\xi},\xi]\Big|_{J=\bar{\xi}=\xi=0},
\end{equation}
by using Eq.(13), we obtain 
\begin{eqnarray}
& &\frac{\partial}{\partial x_{\mu}}\tilde{G}_{rs}^{ab}(x,y)+gf^{acd}\langle T_{C}(A_{r}^{c,\mu}(x)\omega_{r}^{d}(x)\bar{\omega}_{s}^{b}(y))\rangle\nonumber \\
& &+\lambda\frac{\partial}{\partial y^{\underline{\nu}}}G_{rs}^{ab,\mu\underline{\nu}}(x,y)=0.
\end{eqnarray}
Here the gluon-ghost three-point function $\langle
T_{C}(A_{r}^{c,\mu}(x)\omega_{r}^{d}(x)\bar{\omega}_{s}^{b}(y))\rangle
$ is 
\begin{equation}
\langle T_{C}(A_{r}^{c,\mu}(x)\omega_{r}^{d}(x)\bar{\omega}_{s}^{b}(y))\rangle=\frac{\delta\ln\hat{Z}[J,\bar{\xi},\xi]}{i\delta\xi_{s}^{b}(y)i\delta J_{r,\mu}^{c}(x)i\delta\bar{\xi}_{r}^{d}(x)}\Bigg|_{J=\bar{\xi}=\xi=0}
\end{equation}
with no summation over $r$. Note that the third term on the right-hand
side of Eq.(13) does not contribute to Eq.(15). The gluon-ghost three-point function (16) may be written as
\begin{eqnarray}
& & gf^{acd}\langle T_{C}(A_{r}^{c,\mu}(x)\omega_{r}^{d}(x)\bar{\omega}_{s}^{b}(y))\rangle\nonumber \\ 
& &~~~~~=i(-1)^{r}\int d^4w\tilde{\Pi}_{rt}^{ac,\mu}(x,w)\tilde{G}_{ts}^{cb}(w,y).
\end{eqnarray}
$\tilde{\Pi}_{rt}^{ac,\mu}(x,w)$ here is related to a
ghost self-energy part $\tilde{\Pi}_{rt}^{ac}(x,w)$ 
through
\begin{equation}
i\frac{\partial}{\partial x^{\underline{\mu}}}\tilde{\Pi}_{rt}^{ac,\underline{\mu}}(x,w)=\tilde{\Pi}_{rt}^{ac}(x,w).
\end{equation}

From now on we use a bold-face letter to denote a $(8\times 8)$ matrix
in color space, while a caret `\^{}' to denote a $(2\times 2)$ matrix in
``type'' space. For example, a full gluon propagator
$\hat{\boldsymbol{G}}^{\mu\nu}(x,y)$ is a $(8\times 8)$ matrix in color
space with matrix element $\hat{G}^{ab,\mu\nu}(x,y)$, which is a
$(2\times 2)$ matrix in ``type'' space with matrix element
$G_{rs}^{ab,\mu\nu}(x,y)$. Then by using Eq.(17), Eq.(15) can be written as 
\begin{equation}
\frac{\partial}{\partial x_{\mu}}\hat{\tilde{\boldsymbol{G}}}(x,y)-i\hat{\tau}\int d^{4}w\hat{\tilde{\boldsymbol{\Pi}}}^{\mu}(x,w)\hat{\tilde{\boldsymbol{G}}}(w,y)+\lambda\frac{\partial}{\partial y^{\underline{\nu}}}\hat{\boldsymbol{G}}^{\mu\underline{\nu}}(x,y)=0,
\end{equation}
where $\hat{\tau}=$ diag $(1, -1)$.  

We multiply Eq.(19) by the inverse full gluon
propagator $\hat{\boldsymbol{G}}_{\nu\mu}^{-1}(v,x)$ from the left, by the inverse full ghost propagator
$\hat{\tilde{\boldsymbol{G}}}^{-1}(y,z)$ from the right, and then integrate over
$x$ and $y$, to obtain
\begin{equation}
-\frac{\partial}{\partial z_{\mu}}\hat{\boldsymbol{G}}_{\nu\mu}^{-1}(v,z)-i\int d^{4}x\hat{\boldsymbol{G}}_{\nu\mu}^{-1}(v,x)\,\hat{\tau}\,\hat{\tilde{\boldsymbol{\Pi}}}^{\mu}(x,z)-\lambda\frac{\partial}{\partial v^{\underline{\nu}}}\hat{\tilde{\boldsymbol{G}}}^{-1}(v,z)=0.
\end{equation}

Here we recall the Schwinger-Dyson equations:
\begin{eqnarray}
\hat{\boldsymbol{G}}^{\mu\nu}(x,y)&=&\hat{\boldsymbol{\Delta}}^{\mu\nu}(x,y)-i\int d^{4}z\int d^{4}w\hat{\boldsymbol{\Delta}}^{\mu\rho}(x,z)\hat{\boldsymbol{\Pi}}_{\rho\sigma}(z,w)\hat{\boldsymbol{G}}^{\sigma\nu}(w,y)\nonumber\\&=&\hat{\boldsymbol{\Delta}}^{\mu\nu}(x,y)-i\int d^{4}z\int d^{4}w \hat{\boldsymbol{G}}^{\mu\rho}(x,z)\hat{\boldsymbol{\Pi}}_{\rho\sigma}(z,w)\hat{\boldsymbol{\Delta}}^{\sigma\nu}(w,y),\nonumber \\
& &{}\\
\hat{\tilde{\boldsymbol{G}}}(x,y)&=&\hat{\tilde{\boldsymbol{\Delta}}}(x,y)-i\int d^{4}z\int d^{4}w\hat{\tilde{\boldsymbol{\Delta}}}(x,z)\hat{\tilde{\boldsymbol{\Pi}}}(z,w)\hat{\tilde{\boldsymbol{G}}}(w,y)\nonumber\\&=&\hat{\tilde{\boldsymbol{\Delta}}}(x,y)-i\int d^{4}z\int d^{4}w \hat{\tilde{\boldsymbol{G}}}(x,z)\hat{\tilde{\boldsymbol{\Pi}}}(z,w)\hat{\tilde{\boldsymbol{\Delta}}}(w,y),
\end{eqnarray}
where $\hat{\boldsymbol{\Delta}}^{\mu\nu} (\hat{\tilde{\boldsymbol{\Delta}}}) $
is the bare gluon (ghost) propagator. Then we have
\begin{eqnarray}
\hat{\boldsymbol{G}}_{\mu\nu}^{-1}(x,y)&=&\hat{\boldsymbol{\Delta}}_{\mu\nu}^{-1}(x,y)+i\hat{\boldsymbol{\Pi}}_{\mu\nu}(x,y)\nonumber \\
&=&-i\boldsymbol{I}\hat{\tau}\biggl(g_{\mu\nu}\partial_{x}^{2}-\frac{\partial}{\partial x^{\mu}}\frac{\partial}{\partial x^{\nu}}+\lambda\frac{\partial}{\partial x^{\underline{\mu}}}\frac{\partial}{\partial x^{\underline{\nu}}}\biggr)\delta^{4}(x-y)+i\hat{\boldsymbol{\Pi}}_{\mu\nu}(x,y)\nonumber \\
& &{}\\
\hat{\tilde{\boldsymbol{G}}}^{-1}(x,y)&=&\hat{\tilde{\boldsymbol{\Delta}}}^{-1}(x,y)+i\hat{\tilde{\boldsymbol{\Pi}}}(x,y)\nonumber \\
&=&i\boldsymbol{I}\hat{\tau}\nabla_{x}^{2}\delta^{4}(x-y)+i\hat{\tilde{\boldsymbol{\Pi}}}(x,y),
\end{eqnarray}
where $\boldsymbol{I}$ is a $(8\times 8)$ unit matrix in color space.
Substituting Eqs.(23) and (24) into Eq.(20), we finally obtain
\begin{eqnarray}
& &-i\frac{\partial}{\partial z_{\mu}}\hat{\boldsymbol{\Pi}}_{\nu\mu}(v,z)-\biggl(g_{\nu\mu}\partial_{v}^{2}-\frac{\partial}{\partial v^{\nu}}\frac{\partial}{\partial v^{\mu}}\biggr)\hat{\tilde{\boldsymbol{\Pi}}}^{\mu}(v,z)\nonumber \\
& &+\int d^{4}x\hat{\boldsymbol{\Pi}}_{\nu\mu}(v,x)\,\hat{\tau}\,\hat{\tilde{\boldsymbol{\Pi}}}^{\mu}(x,z)=0.
\end{eqnarray}
This is a desired Slavnov-Taylor identity for gluon polarization tensor
in Coulomb gauge. Note that, in the course of derivation, the terms which
explicitly depend on the Coulomb gauge parameter $\lambda$ (Eqs.(20) and
(23)) cancel out with the help of Eq.(18). 

Through similar procedure to the above derivation,
one can derive a covariant-gauge counterpart of Eq.(25). As a matter of
fact, it takes the same form as Eq.(25). 

\section{Out-of-equilibrium QGP}
Here we deal with quasi-uniform QGP near equilibrium or
nonequilibrium quasistationary QGP, which we simply refer to as
out-of-equilibrium QGP. Out-of-equilibrium QGP is characterized
by two different spacetime scales: microscopic or
quantum-field-theoretical and macroscopic or statistical. The first
scale, the microscopic-correlation scale, characterizes the range of
radiative corrections to reactions taking places in the QGP, while
the second scale measures the relaxation of the QGP. A well-known
intuitive picture for dealing with such system is to divide
spacetime into many ``cells'' whose characteristic size,
$L^{\mu}(\mu=0,\cdots, 3)$, is in between microscopic and macroscopic
scales. It is assumed that the correlation between different cells is
negligible in the sense that microscopic or elementary reactions can be
regarded as taking place in a single cell. On the other hand, in a single
cell, relaxation phenomena are negligible.

The above intuitive picture may be implemented as
follows. Let $\Delta(v,z)$ be a generic propagator. For an out-of-equilibrium QGP,
$\Delta(v,z)$, with $v-z$ fixed, does not change
appreciably in the region $|X^{\mu}-X_{0}^{\mu}|\lesssim L^{\mu}$, where
$X^{\mu}\equiv(v^{\mu}+z^{\mu})/2$ is the midpoint and $X_{0}^{\mu}$
is an arbitrary spacetime point. The self-energy part
$\Pi(v,z)$ enjoys a similar property. Thus, $X^{\mu}$ may be used as a label
for the spacetime cells and is called the ``macroscopic spacetime
coordinates.'' On the other hand, relative spacetime coordinates
$v^{\mu}-z^{\mu}$ describe microscopic
reactions taking place in a single spacetime cell. A Fourier
transformation with respect to the relative coordinates
$v^{\mu}-z^{\mu}$ yields
\begin{equation}
\Delta(X;P)\equiv\int d^{4}(v-z)\,e^{iP\cdot(v-z)}\Delta(v,z),
\label{okano} 
\end{equation}
together with a similar formula for $\Pi$. The above
observation shows that $P^{\mu}$ ($\gtrsim 1/L^{\mu}$) in 
Eq.(\ref{okano}) can
be regarded as the momentum of the quasiparticle participating in the
microscopic reaction under consideration. Thus
$\Delta(X;P)$ and
$\Pi(X;P)$ vary slowly in $X$. Then, we employ the derivative expansion,
\begin{eqnarray}
\Delta(X;P)&=&\Biggl[1+(X-Y)^{\sigma}\frac{\partial}{\partial Y^{\sigma}}\nonumber \\
& &+\frac{1}{2!}(X-Y)^{\rho}(X-Y)^{\sigma}\frac{\partial}{\partial Y^{\rho}}\frac{\partial}{\partial Y^{\sigma}}+\cdots\Biggr]\Delta(Y;P),
\end{eqnarray}
together with a similar formula for $\Pi(X;P)$.

Fourier transforming Eq.(25) on $v-z$ and carring out the derivative
expansion, we obtain
\begin{eqnarray}
& & P^{\mu}\hat{\boldsymbol{\Pi}}_{\nu\mu}(X;P)+(g_{\nu\mu}P^{2}-P_{\nu}P_{\mu})\hat{\tilde{\boldsymbol{\Pi}}}^{\mu}(X;P)+\hat{\boldsymbol{\Pi}}_{\nu\mu}(X;P)\,\hat{\tau}\,\hat{\tilde{\boldsymbol{\Pi}}}^{\mu}(X;P)\nonumber \\
& & +\frac{1}{2i}\frac{\partial}{\partial X_{\mu}}\hat{\boldsymbol{\Pi}}_{\nu\mu}(X;P)+\frac{1}{2i}\biggl\{(g_{\nu\mu}P^{2}-P_{\nu}P_{\mu})\,\hat{\tau}+\hat{\boldsymbol{\Pi}}_{\nu\mu},~\hat{\tau}\,\hat{\tilde{\boldsymbol{\Pi}}}^{\mu}\biggr\}_{X,P}\nonumber \\
& &-\frac{1}{4}\biggl(g_{\nu\mu}\partial_{X}^{2}-\frac{\partial}{\partial X^{\nu}}\frac{\partial}{\partial X^{\mu}}\biggr)\hat{\tilde{\boldsymbol{\Pi}}}^{\mu}(X;P) \nonumber \\
& & +\frac{1}{2!}\Bigl(\frac{1}{2i}\Bigr)^{2}\Biggl[\biggl\{\frac{\partial}{\partial X_{\lambda}}\hat{\boldsymbol{\Pi}}_{\nu\mu},~ \hat{\tau}\,\frac{\partial}{\partial P^{\lambda}}\hat{\tilde{\boldsymbol{\Pi}}}^{\mu}\biggr\}_{X,P}-\biggl\{\frac{\partial}{\partial P^{\lambda}}\hat{\boldsymbol{\Pi}}_{\nu\mu},~ \hat{\tau}\,\frac{\partial}{\partial X_{\lambda}}\hat{\tilde{\boldsymbol{\Pi}}}^{\mu}\biggr\}_{X,P}\Biggr]\nonumber \\
& &+O((\partial_{X})^{3})=0,
\end{eqnarray}
where Poisson bracket $\{\cdots,~ \cdots\}_{X,P}$ is defined by
\begin{equation}
\Bigl\{\boldsymbol{\hat{A}}(X;P),~ \boldsymbol{\hat{B}}(X;P)\Bigr\}_{X,P}\equiv \frac{\partial\boldsymbol{\hat{A}}(X;P)}{\partial X_{\mu}}\frac{\partial\boldsymbol{\hat{B}}(X;P)}{\partial P^{\mu}}-\frac{\partial\boldsymbol{\hat{A}}(X;P)}{\partial P^{\mu}}\frac{\partial\boldsymbol{\hat{B}}(X;P)}{\partial X_{\mu}}.
\end{equation}
In view of perturbation theory, 
$\hat{\tilde{\boldsymbol{\Pi}}}^{\mu}$ in Coulomb gauge, Eq.(28), is a diagonal matrix in ``type space''
since the bare ghost propagator $\hat{\tilde{\boldsymbol{\Delta}}}$ is diagonal.

In the following, we restrict our concern to the leading parts of
Eq.(28),  
\begin{equation}
P^{\mu}\hat{\boldsymbol{\Pi}}_{\nu\mu}(X;P)+(g_{\nu\mu}P^{2}-P_{\nu}P_{\mu})\hat{\tilde{\boldsymbol{\Pi}}}^{\mu}(X;P)+\hat{\boldsymbol{\Pi}}_{\nu\mu}\,\hat{\tau}\,\hat{\tilde{\boldsymbol{\Pi}}}^{\mu}(X;P)=0.
\end{equation}
In the case of Coulomb gauge, $\hat{\boldsymbol{\Pi}}_{\nu\mu}(X;P)$ is usually decomposed as
\begin{eqnarray}
\hat{\boldsymbol{\Pi}}_{\nu\mu}(X;P)&=&\mathcal{P}_{\nu\mu}^{T}(\hat{\boldsymbol{p}})\hat{\boldsymbol{\Pi}}^{T}(X;P)+n_{\nu}n_{\mu}\hat{\boldsymbol{\Pi}}^{L}(X;P)\nonumber \\
& &+\frac{p_{0}}{p}\Bigl(\hat{P}_{\underline{\nu}}n_{\mu}+n_{\nu}\hat{P}_{\underline{\mu}}\Bigr)\hat{\boldsymbol{\Pi}}^{C}(X;P)-\hat{P}_{\underline{\nu}}\hat{P}_{\underline{\mu}}\hat{\boldsymbol{\Pi}}^{D}(X;P),
\end{eqnarray}
where $\mathcal{P}_{\nu\mu}^{T}(\hat{\boldsymbol{p}})$ is the transverse
projection operator:
\begin{equation}
\mathcal{P}_{\nu\mu}^{T}(\hat{\boldsymbol{p}})\equiv g_{\underline{\nu}\underline{\mu}}+\hat{P}_{\underline{\nu}}\hat{P}_{\underline{\mu}}.
\end{equation}
Here $g_{\underline{\nu}\underline{\mu}}\equiv -\sum_{i,j=1}^{3}g_{\nu
i}g_{\mu j}\delta^{ij}$ and $\hat{P}^{\underline{\mu}}\equiv
(0,\hat{\boldsymbol{p}})$ with
$\hat{\boldsymbol{p}}\equiv\boldsymbol{p}/p$.

Substituting Eq.(31) into Eq.(30), we obtain 
\begin{eqnarray}
\hat{\boldsymbol{\Pi}}^{L}&=&\hat{\boldsymbol{\Pi}}^{C}+\frac{p}{p_{0}}(pn_{\mu}+p_{0}\hat{P}_{\underline{\mu}})\hat{\tilde{\boldsymbol{\Pi}}}^{\mu}-\biggl(\frac{1}{p_{0}}n_{\mu}\hat{\boldsymbol{\Pi}}^{L}+\frac{1}{p}\hat{P}_{\underline{\mu}}\hat{\boldsymbol{\Pi}}^{C}\biggr)\hat{\tau}\hat{\tilde{\boldsymbol{\Pi}}}^{\mu}, \\
\hat{\boldsymbol{\Pi}}^{L}&=&-\frac{p^{2}}{p_{0}^{2}}\hat{\boldsymbol{\Pi}}^{D}+2\frac{p}{p_{0}}(pn_{\mu}+p_{0}\hat{P}_{\underline{\mu}})\hat{\tilde{\boldsymbol{\Pi}}}^{\mu}+\biggl(-\frac{1}{p_{0}}n_{\mu}\hat{\boldsymbol{\Pi}}^{L}+\frac{p}{p_{0}^{2}}\hat{P}_{\underline{\mu}}\hat{\boldsymbol{\Pi}}^{D}\nonumber \\
& &-\frac{1}{p_{0}}\Bigl(\frac{p_{0}}{p}\hat{P}_{\underline{\mu}}+n_{\mu}\Bigr)\hat{\boldsymbol{\Pi}}^{C}\biggr)\hat{\tau}\hat{\tilde{\boldsymbol{\Pi}}}^{\mu}.
\end{eqnarray}
Note that the identities (33) and (34) are valid to all orders in
perturbation theory. Ward identity in QED plays a key role in the formal discussion of the
theory and simplifies the practical calculation. ST identity derived
here is expected to play equally important role in out-of-equilibrium QCD.

Let $\sum_{n=1}^{\infty}g^{2n}\hat{\boldsymbol{\Pi}}_{\nu\mu}^{(2n)}$
$\bigl[\sum_{n=1}^{\infty}g^{2n}\hat{\tilde{\boldsymbol{\Pi}}}_{\mu}^{(2n)}\bigr]$
be a perturbation series of $\hat{\boldsymbol{\Pi}}_{\nu\mu}$
$\bigl[\hat{\tilde{\boldsymbol{\Pi}}}_{\mu}\bigr]$. Substituting this
into Eqs.(33) and (34), we obtain
\begin{eqnarray}
\hat{\boldsymbol{\Pi}}^{L(2n)}&=&\hat{\boldsymbol{\Pi}}^{C(2n)}+\frac{p}{p_{0}}(pn_{\mu}+p_{0}\hat{P}_{\underline{\mu}})\hat{\tilde{\boldsymbol{\Pi}}}^{(2n)\mu}\nonumber \\
& &-\sum_{m=2}^{2n-2}\biggl(\frac{1}{p_{0}}n_{\mu}\hat{\boldsymbol{\Pi}}^{L(2n-m)}+\frac{1}{p}\hat{P}_{\underline{\mu}}\hat{\boldsymbol{\Pi}}^{C(2n-m)}\biggr)\hat{\tau}\hat{\tilde{\boldsymbol{\Pi}}}^{(m)\mu}, \\
\hat{\boldsymbol{\Pi}}^{L(2n)}&=&-\frac{p^{2}}{p_{0}^{2}}\hat{\boldsymbol{\Pi}}^{D(2n)}+2\frac{p}{p_{0}}(pn_{\mu}+p_{0}\hat{P}_{\underline{\mu}})\hat{\tilde{\boldsymbol{\Pi}}}^{(2n)\mu}+\sum_{m=2}^{2n-2}\biggl(-\frac{1}{p_{0}}n_{\mu}\hat{\boldsymbol{\Pi}}^{L(2n-m)}\nonumber \\
& &+\frac{p}{p_{0}^{2}}\hat{P}_{\underline{\mu}}\hat{\boldsymbol{\Pi}}^{D(2n-m)}-\frac{1}{p_{0}}\Bigl(\frac{p_{0}}{p}\hat{P}_{\underline{\mu}}+n_{\mu}\Bigr)\hat{\boldsymbol{\Pi}}^{C(2n-m)}\biggr)\hat{\tau}\hat{\tilde{\boldsymbol{\Pi}}}^{(m)\mu}.
\end{eqnarray}
Equations (35) and (36) are also valid for the improved perturbation
theory in which the Hard-Thermal-Loops (HTL) resummation is performed
for soft modes [11]. Equations (35) and (36) serve as a consistency
check of the perturbative computation.

Computation of ghost self-energy part is much simpler than that of gluon
self-energy part. This is because the ghost propagator (7) is static and
then diagonal in type space. As an illustration, we compute one-loop
contribution $\hat{\tilde{\boldsymbol{\Pi}}}^{(2)\mu}$. Using Eqs.(7)
and (8), we can write
$\tilde{\Pi}_{rs}^{(2)ab,\mu}$ as
\begin{equation}
-i\tilde{\Pi}_{rs}^{(2)ab,\mu}(X;P)=gf^{ecb}f^{eac}\int\frac{d^{4}K}{(2\pi)^{4}}(P-K)_{\underline{\nu}}\tilde{\Delta}_{rs}(P-K)\Delta_{rs}^{\mu\underline{\nu}}(K)
\end{equation}
with no summation over r and s. In Eq.(37), $\tilde{\Delta}_{rs}$ is as
in Eq.(7), and $\Delta_{11}^{\mu\nu}(K)$
[$\Delta_{22}^{\mu\nu}(K)$] is the (11) [(22)]-component of the gluon propagator:
\begin{eqnarray}
-i\Delta_{11}^{\mu\nu}(K)&=&-i(\Delta_{22}^{\mu\nu}(K))^{*}\nonumber \\
&=&\mathcal{P}^{T,\mu\nu}(\hat{\boldsymbol{k}})\frac{-1}{K^{2}+i0^{+}}+\frac{n^{\mu}n^{\nu}}{k^{2}}-\frac{1}{\lambda}\frac{K^{\mu}K^{\nu}}{k^{4}}\nonumber \\
& &+i\mathcal{P}^{T,\mu\nu}(\hat{\boldsymbol{k}})2\pi\delta(K^{2})\Bigl(\theta(k_{0})n(X;k,\hat{\boldsymbol{k}})+\theta(-k_{0})n(X;k,-\hat{\boldsymbol{k}})\Bigr).\nonumber \\
& &
\end{eqnarray}
Here $n(X;k,\hat{\boldsymbol{k}})$
is the number density of the gluon with momentum $\boldsymbol{k}$ at the
spacetime point $X^{\sigma}$. From Eq.(37) with Eq.(38), one can readily
see that
$\tilde{\Pi}_{12}^{(2)ab,\mu}=\tilde{\Pi}_{21}^{(2)ab,\mu}=0$ and
$(\tilde{\Pi}_{22}^{(2)ab,\mu})^{*}=-\tilde{\Pi}_{11}^{(2)ab,\mu}$. Straightforward manipulation yields
\begin{eqnarray}
\hat{\tilde{\boldsymbol{\Pi}}}^{(2)\mu=0}&=&0,\\
\tilde{\boldsymbol{\Pi}}_{11}^{(2)\underline{\mu}}(X;P)&=&\frac{3}{2}g^{2}\boldsymbol{I}\int \frac{d^{3}k}{(2\pi)^{3}}\frac{-P^{\underline{\mu}}+(\boldsymbol{p}\cdot\boldsymbol{\hat{k}})\hat{K}^{\underline{\mu}}}{k(\boldsymbol{p}-\boldsymbol{k})^{2}}\Bigl(n(X;k,\hat{\boldsymbol{k}})+n(X;k,-\hat{\boldsymbol{k}})\Bigr)\nonumber \\
& &+\cdots,
\end{eqnarray}
where `$\cdots$' stands for the contribution from vacuum theory, which
depends on the renormalization scheme. For the isotropic QGP, $n(X;k,\hat{\boldsymbol{k}})=n(X;k)$, Eq.(40) turns out to
\begin{eqnarray} 
\tilde{\boldsymbol{\Pi}}_{11}^{(2)\underline{\mu}}(X;P)&=&-3\pi g^{2}\boldsymbol{I}\hat{P}^{\underline{\mu}}\int_{0}^{\infty}\frac{dk}{(2\pi)^{3}}n(X;k)\biggl[\frac{p^{2}+k^{2}}{pk}+\frac{(p^{2}-k^{2})^{2}}{2p^{2}k^{2}}\ln\frac{p-k}{p+k}\biggr]\nonumber \\
& &+\cdots.
\end{eqnarray}
Note that $\tilde{\boldsymbol{\Pi}}_{11}^{(2)\underline{\mu}}$ in
Eqs.(40) and (41) are independent of $p_{0}$. Thus, Eqs.(35) and (36)
tell us that for obtaing the components of the gluon self-energy parts,
$\hat{\boldsymbol{\Pi}}^{P(2)} (P=L,C,D)$, computation of one of them is
sufficient. It is to be noted that the imaginary part of
$\hat{\boldsymbol{\Pi}}^{L}$ on the mass-shell is proportional to the damping rate for
longitudinal gluon, which is a propagating mode of gluonic
quasiparticle (plasmon) in QGP. 

In the case of covariant gauge, $\hat{\boldsymbol{\Pi}}_{\nu\mu}(X;P)$
is usually decomposed as [6] 
\begin{eqnarray}
\hat{\boldsymbol{\Pi}}_{\nu\mu}(X;P)&=&\mathcal{P}_{\nu\mu}^{T}(\hat{\boldsymbol{p}})\hat{\boldsymbol{\Pi}}^{T}(X;P)+\mathcal{P}_{\nu\mu}^{L}(P)\hat{\boldsymbol{\Pi}}^{L}(X;P)\nonumber \\
& &+\mathcal{C}_{\nu\mu}(P)\hat{\boldsymbol{\Pi}}^{C}(X;P)+\mathcal{D}_{\nu\mu}(P)\hat{\boldsymbol{\Pi}}^{D}(X;P),
\end{eqnarray}
where, $\mathcal{P}_{\nu\mu}^{T}(\hat{\boldsymbol{p}})$ is as in Eq.(32) and
\begin{eqnarray}
\mathcal{P}_{\nu\mu}^{L}(P)&\equiv& g_{\nu\mu}-\frac{P_{\nu}P_{\mu}}{P^{2}+i0^{+}}-\mathcal{P}_{\nu\mu}^{T}(\hat{\boldsymbol{p}}),\\
\mathcal{C}_{\nu\mu}(P)&\equiv&\frac{1}{\sqrt{2}\,p_{0}p}\biggl(P_{\nu}P_{\underline{\mu}}+P_{\underline{\nu}}P_{\mu}+2p^{2}\frac{P_{\nu}P_{\mu}}{P^{2}+i0^{+}}\biggr),\\
\mathcal{D}_{\nu\mu}(P)&\equiv&\frac{P_{\nu}P_{\mu}}{P^{2}+i0^{+}}.
\end{eqnarray}
Note that $\hat{\boldsymbol{\Pi}}^{T}$, $\hat{\boldsymbol{\Pi}}^{L}$,
$\hat{\boldsymbol{\Pi}}^{C}$ and $\hat{\boldsymbol{\Pi}}^{D}$ in Eq.(42)
are different from those in Eq.(31).

Substituting Eq.(42) into Eq.(30), we obtain the covariant-gauge
counterparts of Eqs.(33) and (34):
\begin{eqnarray}
\hat{\boldsymbol{\Pi}}^{C}&=&\sqrt{2}(pn_{\mu}+p_{0}\hat{P}_{\underline{\mu}})\hat{\tilde{\boldsymbol{\Pi}}}^{\mu}-\biggl(\frac{\sqrt{2}}{p}\mathcal{P}_{0\mu}^{L}\hat{\boldsymbol{\Pi}}^{L}+\frac{P_{\mu}}{P^{2}+i0^{+}}\hat{\boldsymbol{\Pi}}^{C}\biggr)\hat{\tau}\hat{\tilde{\boldsymbol{\Pi}}}^{\mu},\\
\hat{\boldsymbol{\Pi}}^{D}&=&\frac{-1}{P^{2}+i0^{+}}\biggl(\frac{pn_{\mu}+p_{0}\hat{P}_{\underline{\mu}}}{\sqrt{2}}\hat{\boldsymbol{\Pi}}^{C}+P_{\mu}\hat{\boldsymbol{\Pi}}^{D}\biggr)\hat{\tau}\hat{\tilde{\boldsymbol{\Pi}}}^{\mu}.
\end{eqnarray}
Substituting the perturbation series of
$\hat{\boldsymbol{\Pi}}_{\nu\mu}$ and
$\hat{\tilde{\boldsymbol{\Pi}}}_{\mu}$ into Eqs.(46) and (47), we obtain
\begin{eqnarray}
\hat{\boldsymbol{\Pi}}^{C(2n)}&=&\sqrt{2}(pn_{\mu}+p_{0}\hat{P}_{\underline{\mu}})\hat{\tilde{\boldsymbol{\Pi}}}^{(2n)\mu}\nonumber \\
& &-\sum_{m=2}^{2n-2}\biggl(\frac{\sqrt{2}}{p}\mathcal{P}_{0\mu}^{L}\hat{\boldsymbol{\Pi}}^{L(2n-m)}+\frac{P_{\mu}}{P^{2}+i0^{+}}\hat{\boldsymbol{\Pi}}^{C(2n-m)}\biggr)\hat{\tau}\hat{\tilde{\boldsymbol{\Pi}}}^{(m)\mu},\\
\hat{\boldsymbol{\Pi}}^{D(2n)}&=&\frac{-1}{P^{2}+i0^{+}}\sum_{m=2}^{2n-2}\biggl(\frac{pn_{\mu}+p_{0}\hat{P}_{\underline{\mu}}}{\sqrt{2}}\hat{\boldsymbol{\Pi}}^{C(2n-m)}+P_{\mu}\hat{\boldsymbol{\Pi}}^{D(2n-m)}\biggr)\hat{\tau}\hat{\tilde{\boldsymbol{\Pi}}}^{(m)\mu}.\nonumber \\
& &
\end{eqnarray}
Some observations are in order.
\begin{description}
\item[1.] $\hat{\boldsymbol{\Pi}}^{C(2)}=\sqrt{2}(pn_{\mu}+p_{0}\hat{P}_{\underline{\mu}})\hat{\tilde{\boldsymbol{\Pi}}}^{(2)\mu}$
and $\hat{\boldsymbol{\Pi}}^{D(2)}=0$. As in the case of Coulomb gauge, computation of
$\hat{\tilde{\boldsymbol{\Pi}}}^{(2)\mu}$ is relatively easy.
\item[2.] $\hat{\boldsymbol{\Pi}}^{C(2n)}$ is written in terms of
$\hat{\tilde{\boldsymbol{\Pi}}}^{(2m)\mu}$ and
$\hat{\boldsymbol{\Pi}}^{L(2m-2)}$ with $m\leq n$.
\item[3.] $\hat{\boldsymbol{\Pi}}^{D(2n)}$ starts from
$\hat{\boldsymbol{\Pi}}^{D(4)}$ and is written in terms of
$\hat{\tilde{\boldsymbol{\Pi}}}^{(2m)\mu}$ and
$\hat{\boldsymbol{\Pi}}^{L(2m-2)}$ with $m\leq n-1$.
\end{description}

\section{Acknowledgements}
I would like to thank Prof. A. Ni\'{e}gawa for helpful discussions and
careful reading of the manuscript.

\end{document}